\documentclass[10pt,letterpaper,twocolumn]{article} %% two column, final layout

\usepackage{ol2}
\usepackage[draft]{hyperref}
\usepackage{amsmath}

\begin{document}

\twocolumn[ %% activate for two-column option

\title{Non-Hermitian gauged topological laser arrays}

%% For REVTeX it is possible to automate superscript and e-mail callouts with the superscriptaddress option; see REVTeX4 documentation.

\author{Stefano Longhi}

\address{Dipartimento di Fisica, Politecnico di Milano and Istituto di Fotonica e Nanotecnologie del Consiglio Nazionale delle Ricerche, Piazza L. da Vinci 32, I-20133 Milano, Italy (stefano.longhi@polimi.it)}

\begin{abstract}
Stable and phase-locked emission in an extended topological supermode of coupled laser arrays, based on concepts of non-Hermitian and topological photonics, is theoretically suggested. We consider a non-Hermitian Su-Schrieffer-Heeger chain of coupled microring resonators  and show that application of a synthetic imaginary gauge field via auxiliary passive microrings leads to all supermodes of the chain, except one, to become edge states. The only extended supermode, that retains some topological protection, can stably oscillate suppressing all other non-topological edge supermodes. Numerical simulations based on a rate equation model of semiconductor  laser arrays confirm stable anti-phase laser emission in the extended topological supermode and the role of the synthetic gauge field to enhance laser stability
\end{abstract}
 ] %% activate for two-column option

{\it Introduction}. Non-Hermiticity, topology and synthetic gauge fields are important physical concepts that are catching a huge attention in photonics (see \cite{r1,r1bis,r2,r3,r4,r5,r5bis} and references therein). Such concepts have been proven to be very fruitful, among others, when applied to integrated laser devices. For example, recent experiments have demonstrated single-mode microlaser operation based on the concepts of parity-time ($\mathcal{PT}$) symmetry and exceptional points \cite{r6,r7,r8,r9,r10}, simultaneous lasing and anti-lasing \cite{r11}, and topological lasers in photonic crystals \cite{r12,r13}. The interplay between topology and non-Hermiticity is a promising research arena \cite{r14,r15,r16,r17,r18,r20,r21} which is expected to provide further advances in both theoretical and applied aspects of integrated photonics. Recent experiments  \cite{r23,r24} demonstrated lasing of topologically-protected edge states in a non-Hermitian version of the celebrated two-band Su-Schrieffer-Heeger (SSH) model \cite{r14,r16,r18,r22}. In these experiments stable laser operation is observed in the exponentially-localized topologically-protected zero energy mode of a SSH chain of coupled semiconductor microring resonators, while all other extended supermodes of the array are effectively suppressed via selective non-Hermitian ($\mathcal{PT}$-symmetric) pumping configuration \cite{r16}. The lasing mode retains some topological protection of the conservative SSH model, such as robustness against perturbations of coupling constants that do not close the gap. However, a main limitation of such a topological mode is that it is exponentially localized in few cavities of the array, i.e. it does not exploit efficiently the gain available in the chain and is not suited to realize broad-area high-power emission from the laser array. Also, emission in the topological edge mode can be destabilized via a complex phase transition \cite{r24}. On the other hand, it is known that oscillation of a spatially-extended supermode in a laser array -- a longstanding problem in laser science and technology \cite{r25,r26}-- is often prevented by supermode competition, phase unlocking and other laser instabilities \cite{r25,r27,r28,r29,r30}, requiring a careful laser design such as gain tailoring \cite{r31} or the use of diffractive or far-field coupling \cite{r32,r33,r34} to achieve phase locking.\par 
In this Letter I suggest a route toward stable laser emission in a topological and {\it spatially-extended} supermode of an array of coupled microresonators, which is inspired by the concepts of  topology, non-Hermiticity and synthetic gauge fields. I consider a SSH array of coupled microring resonators in a $\mathcal{PT}$-symmetric pumping configuration, which sustains a topologically-protected edge state \cite{r23,r24}, and introduce an artificial imaginary gauge field \cite{r35} via non-Hermitian resonator coupling engineering \cite{r36}. The gauge field transforms the edge state into an extended state and all other extended supermodes of the array into non-topological edge states. In other words, the synthetic imaginary gauge field  realizes an intriguing engineering of laser array supermodes,  squeezing all extended supermodes of the SSH chain at one edge while making the topologically-protected state extended over the array. In this way, the entire gain of the array is efficiently exploited and broad area phase-locked laser emission with a superior stability can be achieved.\par
 {\it Topological laser array with a synthetic imaginary gauge field: supermode analysis}.
  We consider a chain of active/passive evanescently-coupled microrings shown in Fig.1(a), which realizes a non-Hermitian extension of the SSH 
model \cite{r14,r15,r18,r23,r24}. The chain comprises $(2N_d+1)$ microrings, with $(N_d+1)$ active (pumped) rings with optical gain $g_A$ (sublattice A) and $N_d$ passive (dissipative) rings with optical loss $-g_B$ (sublattice B). The microrings have the same resonance frequency and are evanescently-coupled with alternating coupling constants $t_1$ and $t_2>t_1$. The chain is terminated by cells A. A synthetic imaginary gauge field $h$ (complex Peierls$^{\prime}$  phase \cite{r34})  is introduced for coupling constants using auxiliary passive microrings in an antiresonance configuration \cite{r36} that indirectly couple sites A and B [Fig.1(b)], as briefly discussed below.
Indicating by $E^{(A)}_{n}$, $E^{(B)}_{n}$ the modal amplitudes of the fields circulating in the $n$-th microrings of sublattices A and B of the chain, coupled-mode equations of the non-Hermitian SSH lattice read 
 \begin{figure}[htb]
\centerline{\includegraphics[width=8.4cm]{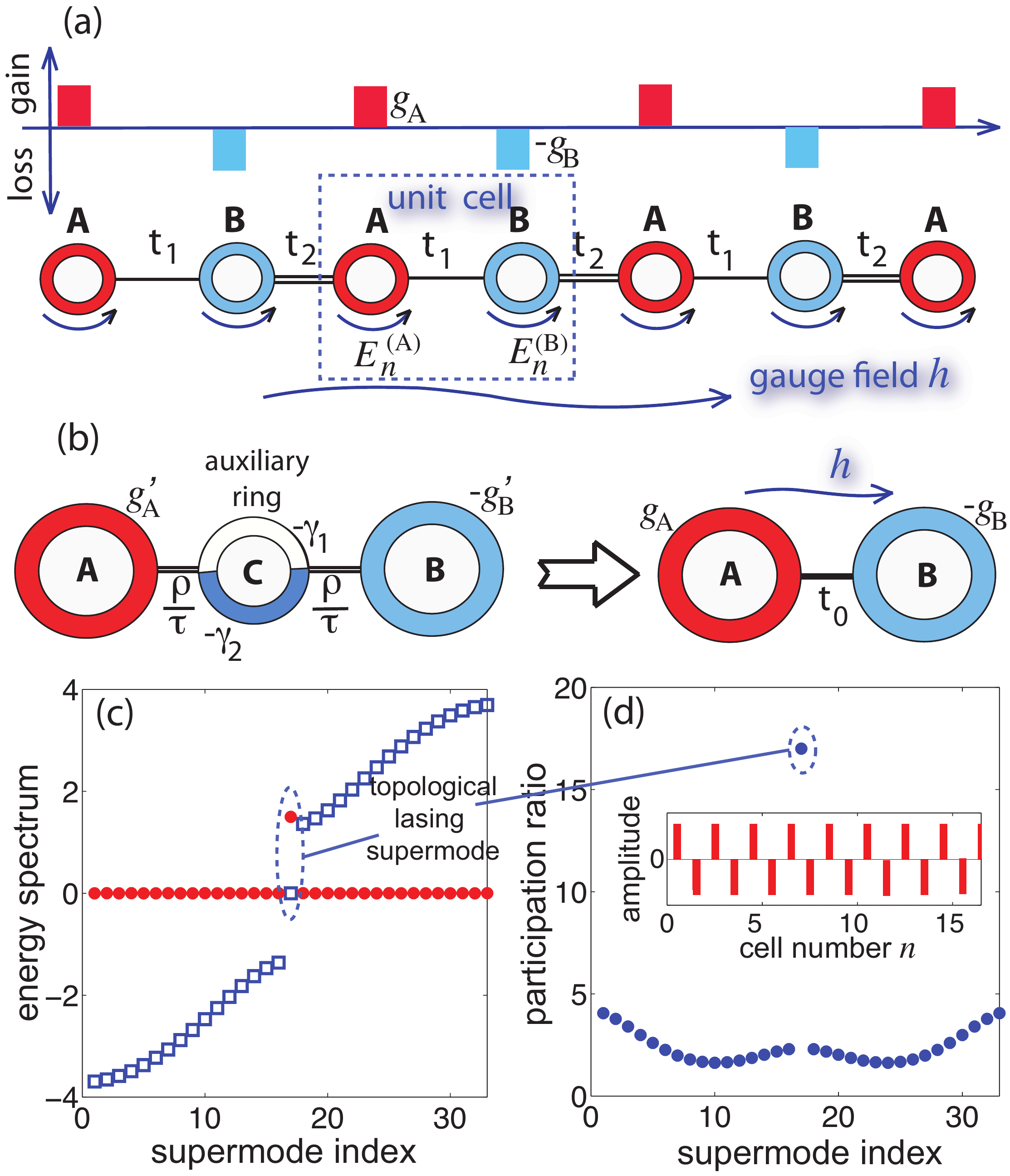}} \caption{ \small
(Color online) (a) Schematic of a non-Hermitian SSH lattice made of an array of $(2Nd+1)$ coupled microring resonators with alternating gain and loss. (b) A synthetic imaginary gauge field between two microrings A and B is obtained by use of an auxiliary ring C in antiresonance with the main rings. The auxiliary ring provides indirect coupling of the electric modal fields in rings A and B with an effective coupling constant $t_0$ and imaginary gauge field $h$ given by Eq.(5). (c) Energy spectrum in units of $t_1$ (open squares: real part; filled circles: imaginary part) of the $\mathcal{PT}$-symmetric SSH Hamiltonian comprising $(2N_d+1)=33$ microrings for parameter values $t_2/t_1=3$ and $g/t_1=1.5$.  The SSH lattice is in phase I, corresponding to real energies of all extended Bloch modes and a single supermode with non vanishing imaginary part  (topological lasing mode). The energy spectrum is independent of the gauge field $h$, whereas the eigenvectors (array supermodes) depend on $h$. (d) Participation ratio of eigenvectors of the SSH lattice with an imaginary gauge field $h= -0.5493$ satisfying Eq.(3). The extended supermode, corresponding to the largest value of PR, is the topological supermode with anti-phase field amplitude distribution shown in the inset.}
\end{figure} 
  \begin{figure}[htb]
\centerline{\includegraphics[width=8.6cm]{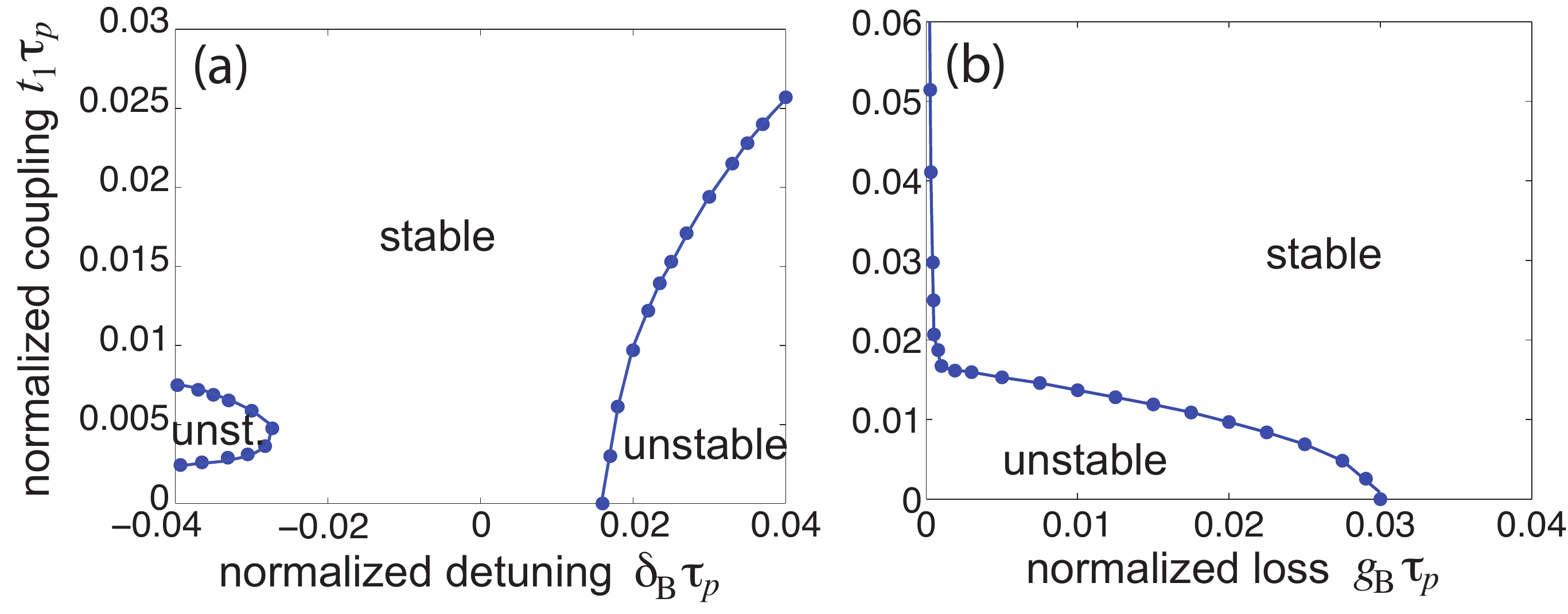}} \caption{ \small
(Color online) Stability diagrams of the topological supermode for non-vanishing cavity detuning $\delta_B$ for parameter values $\alpha=3$, $p_A=0.02$, $t_2/t_1=3$, $h=-0.5493$, and $\tau_s/ \tau_p=2 \times 10^3$. In (a) $g_B \tau_p=0.05$ while in (b) $\delta_B \tau_p=0.01$.}
\end{figure}  
\begin{eqnarray}
i (dE^{(A)}_n/dt) = i g_A E^{(A)}_n+ t_1 \exp(h) E^{(B)}_n \nonumber \\
+ t_2 \exp(-h)E^{(B)}_{n-1} \\
i (dE^{(B)}_n/dt) = -i g_B E^{(B)}_n+ t_1 \exp(-h) E^{(A)}_n \nonumber \\
+  t_2 \exp(h)E^{(A)}_{n+1}  
\end{eqnarray}
with $E^{(B)}_{0}=E^{(B)}_{N_d+1}=0$. A $\mathcal{PT}$-symmetric pumping configuration corresponds to $g_B=g_A \equiv g$. In this case, for $h=0$ the energy spectrum of the SSH Hamiltonian $\mathcal{H}$, defined by Eqs.(1) and (2), comprises the topological edge state $E^{(A)}_n=(-t_1/t_2)^n$, $E^{(B)}_n=0$ with complex energy $\mathcal{E}=ig$, exponentially localized at the left edge of the lattice, and $2N_d$  extended supermodes with energies $\mathcal{E}_{\pm}(q)=\pm \sqrt{t_1^2+t_2^2+2 t_1 t_2 \cos q-g^2}$, where $q=2 l \pi/(N_d+1)$ ($l=1,2,...,N_d$) is the Bloch wave number (it is quantized owing to open boundary conditions). Depending on the value of $g$, the SSH lattice can be found in three different phases \cite{r18,r24}. For $g< (t_2-t_1)$, the lattice is in phase I, corresponding to the unbroken $\mathcal{PT}$ phase (real energies) of all extended supermodes [Fig.1(c)]. This pumping configuration forces the topological edge mode  to lase, while suppressing all other extended modes \cite{r23,r24}.\\ 
In a lattice with open boundary conditions, a non-vanishing imaginary gauge field $h$ does not change the energy spectrum of $\mathcal{H}$, while it changes the localization properties of its eigenvectors \cite{r35,r37}. 
In fact, the imaginary field $h$ can be removed after the imaginary gauge transformation \cite{r35} $E_n^{(A)}=a_n \exp(-2hn)$ and $E_n^{(B)}=b_n \exp(-2hn-h)$, so that $a_n$ and $b_n$ satisfy Eqs.(1) and (2) with $h=0$. Note that, while the energy spectrum of $\mathcal{H}$ remains unchanged, the gauge transformation alters the localization properties of supermodes, which are squeezed toward the left edge (for $h>0$) or the right edge (for $h<0$) of the chain. In particular, if we tune the gauge field $h$ to the negative value 
\begin{equation}
h=-(1/2) {\rm log}(t_2/t_1),
\end{equation}
 it follows that the eigenvector corresponding to the topological edge mode becomes an extended mode with balanced excitation of sites in sublattice A and zero excitation of sublattice B [inset in Fig.1(d)], while all other extended supermodes of $\mathcal{H}$ are squeezed toward the right edge of the chain, i.e. they become (non-topological) right edge states. This is shown in Fig.1(d), where the participation ratio ${\rm PR}=(\sum_n (|E^{(A)}_{n}|^2+|E^{(B)}_n|^2)^2 / \sum_n (|E^{(A)}_{n}|^4+|E^{(B)}_n|^4)$ of various supermodes is depicted. A localized supermode corresponds to a small value of the PR, while an extended supermode shows a large value of the PR. Note that all modes, except one, are localized. The only extended supermode is the topological supermode. Since the gauge transformation does not change the energy spectrum of $\mathcal{H}$, it follows that the SSH lattice with properly-tailored imaginary gauge field can stably lase in the extended topological supermode, in which all active (pumped) microrings are lasing in anti-phase [inset of Fig.1(d)]. Therefore phase-locked broad-area laser emission in a topological supermode is realized.\\ 
 %The extended supermode retains some topological protection of the original non-trivial edge state of the SSH model, i.e. it is robust against disorder in the coupling constants $t_1$ and $t_2$ that do not close the gap of the lattice minibands.\\
 Finally, let us briefly discuss how to realize a synthetic imaginary gauge field. Following a recent proposal \cite{r36}, let us consider the trimer shown in Fig.1(b), in which two main microrings A and B with optical gain/loss $g^{\prime}_A$ and $-g^{\prime}_B$ are indirectly coupled via an auxiliary microring C. The auxiliary ring C is designed to be antiresonant to the main rings A and B, i.e., the length of the auxiliary ring is slightly larger (or smaller) than the main rings so as the field acquires an extra $\pi$ phase shift at each round trip. The auxiliary ring is assumed to be purely passive, providing asymmetric attenuation of the light field in the upper and lower half perimeter, with single-pass losses $-\gamma_1$  and $-\gamma_2$, respectively [Fig.1(b)]. In the mean-field limit, the field amplitude in the auxiliary ring can be eliminated following a procedure similar to the one detailed in Ref.\cite{r36}, yielding the following effective coupled-mode equations for the field amplitudes $E^{(A)}$ and $E^{(B)}$ in the main rings 
 \begin{equation}
 i (dE^{(A,B)}/dt)=t_0 \exp(\pm h) E^{(B,A)} \pm i g_{A,B}E^{(A,B)}
 \end{equation}
 where we have set $g_A \equiv g^{\prime}_A-\sigma$, $g_B \equiv g^{\prime}_B+\sigma$, $\sigma \equiv  (\rho^2 / 2 \tau) {\rm tanh} [(\gamma_1+\gamma_2)/2]$, $\tau$ is the round-trip time in the main rings, $ \rho / \tau$ is the direct coupling  constant between main rings and the auxiliary ring ($\rho \ll 1$), and $t_0$, $h$ are the effective coupling constant and imaginary gauge field, given by
 \begin{equation}
t_0 \equiv  \frac{\rho^2}{\sqrt{2} \tau} \frac{1}{\sqrt{1+ {\rm cosh}(\gamma_1+\gamma_2)}} \;, \;\; h \equiv \frac{\gamma_2-\gamma_1}{2}.
  \end{equation}
  A negative gauge field can be realized by assuming $\gamma_1>\gamma_2$, with $\gamma_2=0$ in the configuration with lowest extra loss term $\sigma$.
  \par
 {\it Semiconductor laser rate equation analysis.}  In a semiconductor laser array, dynamical instabilities are known to arise from supermode competition and from nonlinear-induced resonance detuning of coupled cavities that can prevent phase locking \cite{r28,r29,r30}. Therefore, a proper rate equation model that accounts for nonlinear coupled dynamics of electric modal fields and carrier densities in the active (pumped) microrings of sublattice A should be considered. In the following analysis, resonators of sublattice B are considered  as linear passive cavities with dominant non-resonant (i.e. carrier-independent) optical losses. Assuming that each microring oscillates in a single longitudinal and transverse mode with the same resonance frequency in each sublattice, the laser rate equations in dimensionless form read \cite{r29,r30}
 \begin{eqnarray}
 \frac{1}{\tau_p} \frac{dE^{(A)}_n}{d \tau}  =  (1-i \alpha) Z^{(A)}_n E^{(A)}_n / \tau_p  - it_1 \exp(h) E^{(B)}_{n} \nonumber \\
   -  i t_2 \exp(- h) E^{(B)}_{n -1}  \\
  \frac{1}{\tau_p} \frac{dE^{(B)}_n}{d \tau}  =  -(g_B-i \delta_B)  E^{(B)}_n  - it_1 \exp(-h) E^{(A)}_{n} \nonumber \\
  - i t_2 \exp( h) E^{(A)}_{n +1} \\ 
 \frac{\tau_s}{ \tau_p} \frac{d Z^{(A)}_n }{d \tau}  = p_A-Z_n^{(A)}-(1+2Z_n^{(A)})|E_n^{(A)}|^2. 
  \end{eqnarray}
  \begin{figure}[htb]
\centerline{\includegraphics[width=8.6cm]{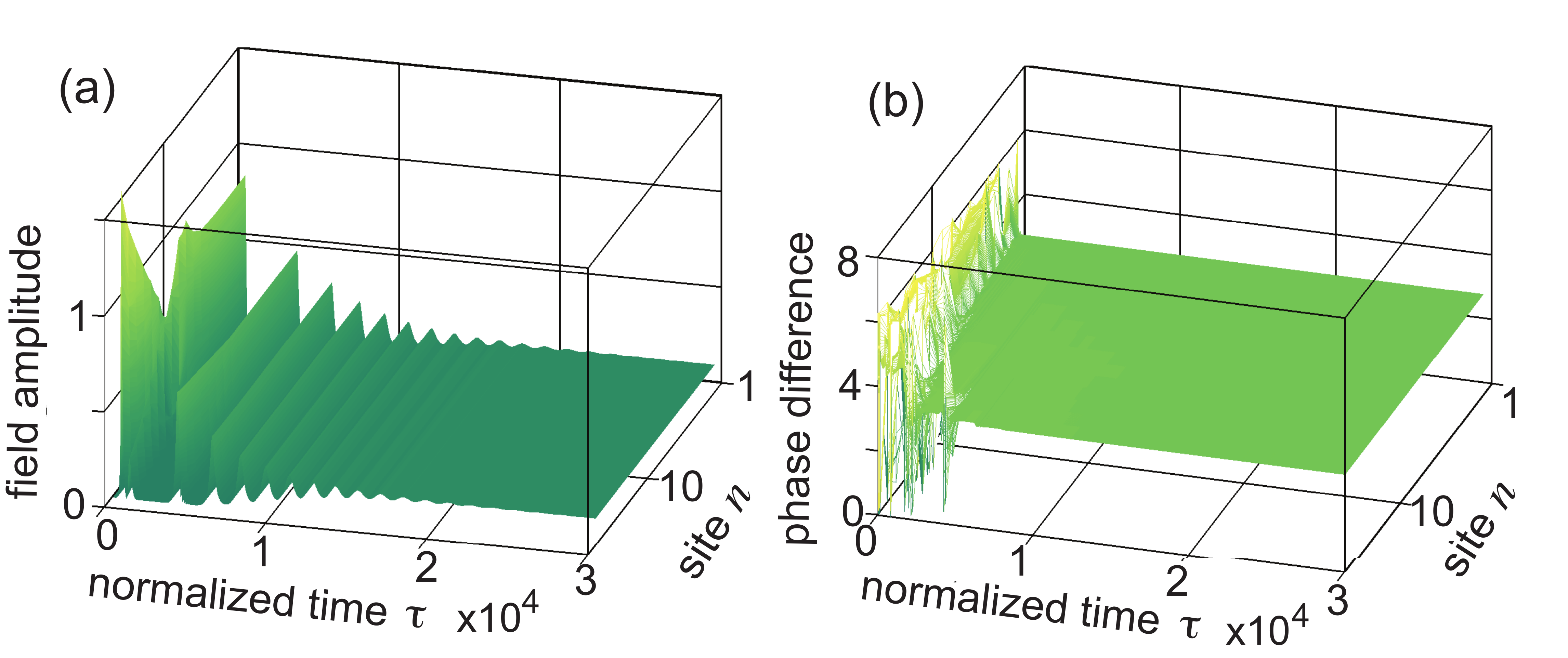}} \caption{ \small
(Color online) Numerical simulations showing lasing emission in the topological extended supermode starting from random small amplitudes of modes $E_n^{(A,B)}$ and equilibrium carrier density $Z_n^{(A)}=p_A$. Parameter values are: $p_A=g_A \tau_p=0.02$, $g_B \tau_p=0.02$, $\delta_B \tau_p=0$, $\tau_s / \tau_p= 2 \times 10^3$,  $\alpha=3$, $ t_1 \tau_p=0.02$, $t_2/t_1=3$ and $h=-0.5493$. (a) Temporal behavior of normalized field amplitudes $|E_n^{(A)}|$ of sublattice A and (b) temporal behavior of phase difference between adjacent modes $\Delta \varphi_n=\varphi_{n+1}-\varphi_n$, where $E_n^{(A)}(\tau)=|E_n^{(A)}(\tau)| \exp [i \varphi_n (\tau)]$. Note that, after an initial relaxation oscillations transient, the array emits in the extended topological supermode with homogeneous amplitudes $|E_n^{(A)}|=\sqrt{p_A}$  in anti-phase ($\Delta \varphi_n= \pi$). The amplitudes of modes in subattice B, not shown in the figure, remain small and are damped toward zero.}
\end{figure} 
\begin{figure}[htb]
\centerline{\includegraphics[width=8.6cm]{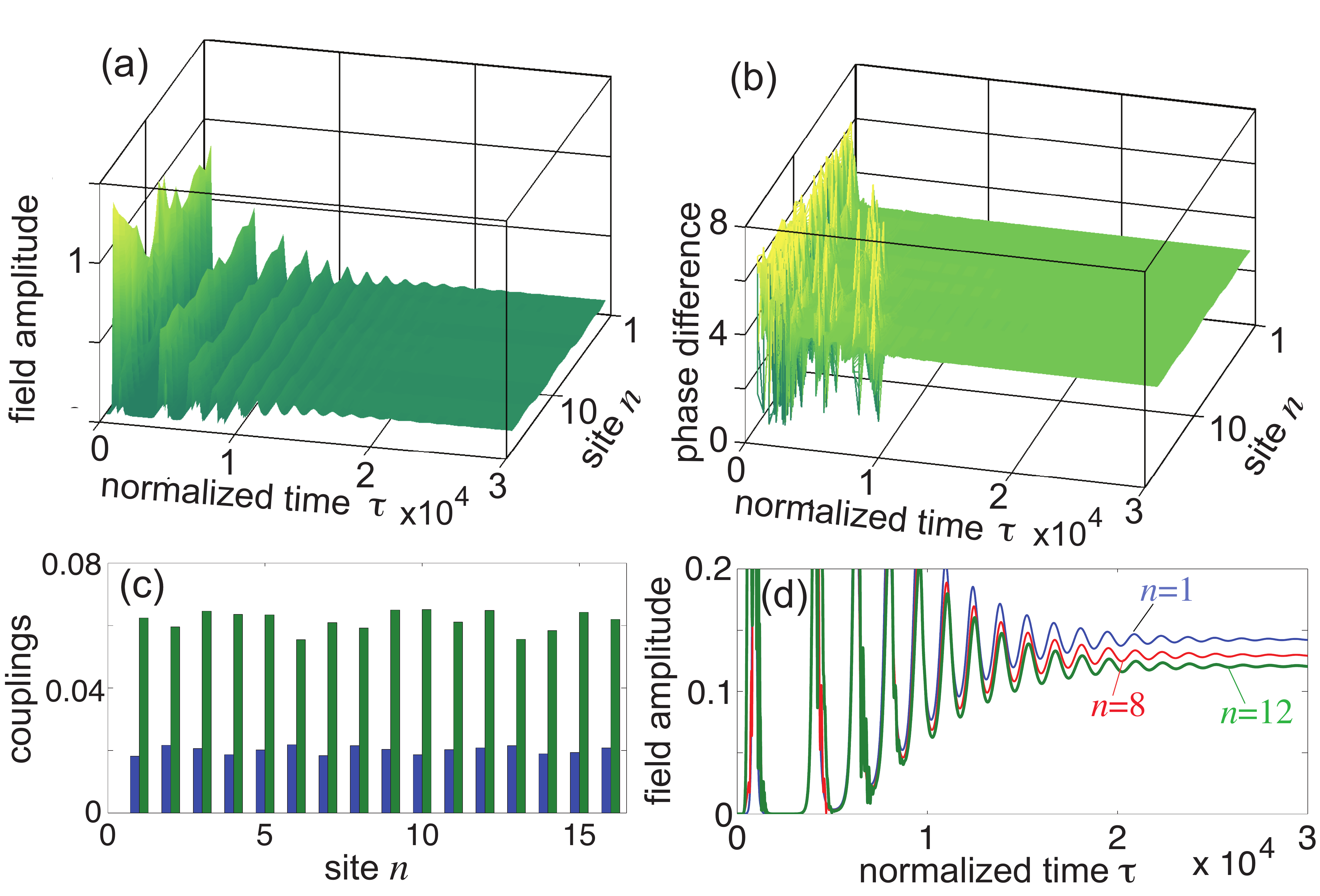}} \caption{ \small
(Color online) (a,b) Same as Fig.3, but with inhomogeneous coupling constants obtained from the mean values $t_1 \tau_p=0.02$ and $t_2 \tau_p=0.06$ adding some perturbations which do not close the gap. The behavior of normalized coupling constants is shown in panel (c) ( $t_1 \tau_p$: lowers bars; $t_2 \tau_p$: higher bars). 
Panel (d) shows the detailed transient built up of modal amplitudes $|E_{n}^{(A)} (\tau)|$ in the three sites $n=1,8,12$ of sublattice A.}
\end{figure}  
 In the above equations, $E^{(A,B)}_n$ are the normalized electric fields in the rings of the two sublattices A and B, $Z_n^{(A)}$ is the normalized excess carrier density in the active rings of sublattice A, $\tau=t/ \tau_p$ is the time variable normalized to the photon lifetime $\tau_p$ of microrings A, $\tau_s$ is the spontaneous carrier lifetime, $\alpha$ is the linewidth enhancement
factor, and $p_A$ is the normalized excess pump current in the active rings (providing a linear gain $g_A=p_A/ \tau_p$). In writing Eqs.(6-8), we also allowed rather generally for a detuning $\delta_B$ of the resonance frequency of modes in the two sublattices. For the imaginary gauge field $h$ satisfying Eq.(3), a steady-state solution to Eqs.(6-8), corresponding to the extended topological supermode shown in the inset of Fig.1(d), is given by $E_n^{(A)}=(-1)^{n} \sqrt{p_A}$, $E_{n}^{(B)}=0$ and $Z_n^{(A)}=0$. Stability of the stationary topological supermode can be investigated by standard linear stability analysis \cite{r28,r30}. The growth rate of perturbations can be determined from the roots of a fifth-order algebraic equation. The coefficients of the polynomial depend on a real parameter $q$, the Bloch wave number of perturbations, which  is quantized owing to the open boundary condition of the chain \cite{r30}. For enough long chains, $q$ can be taken as an almost continuous variable and the stability boundary is rather insensitive to the chain length, the most unstable perturbation being the one corresponding to the wave number $q= \pi$; technical details of the linear stability analysis will be given elsewhere. Numerical results show that the topological supermode is always a locally stable state when $\delta_B=0$ and $g_B>0$, i.e. for resonance condition of cavities in sublattices A and B and for some loss in the passive cavities. On the other hand, instabilities can arise in the detuned case $\delta_B \neq 0$, with the major critical case being the detuning side with $\delta_B \alpha >0$; in particular for $\delta_B \neq 0$ and $g_B=0$ the topological mode is always linearly unstable. Examples of numerically-computed stability boundaries of the topological supermode for detuned operation are shown in Fig.2. Note that, for a non vanishing resonance detuning $\delta_B$, stability of the topological supermode generally requires a minimum threshold value of the coupling [Fig.2(a)] or large enough dissipation in cavities B [Fig.2(b)]. Figures 3 and 4 show examples of laser built up and stable oscillation in the topological extended supermode, after relaxation oscillation transient, as obtained from direct numerical simulations of Eqs.(6-8) for resonance operation $\delta_B=0$. Initial condition is a small random noise of field amplitudes and equilibrium carrier densities. The extended supermode retains some topological protection of the original Hermitian SSH model, i.e. the state is robust against moderate disorder in the coupling constants $t_1$ and $t_2$ that does not close the gap, as shown in Fig.4. In this case the amplitudes $E_{n}^{(A)}$ oscillate in anti-phase, like in the ordered structure, while their intensities reach steady-state but inhomogeneous as a result of disorder in the coupling constants [Fig.4(d)]. Finally, it is worth comparing the behavior of the SSH laser array with and without the synthetic gauge field. Figure 5 shows the transient built up of the same laser array of Fig.3, but with $h=0$. Clearly, in this case the laser emission is very irregular and does not occur on the edge-state topological supermode. Therefore, the synthetic gauge field enhances topological mode stability. \par
 {\it Conclusion.} Stable and broad-area emission in a SSH laser array supporting an extended topological supermode has been theoretically suggested. The laser design exploits concepts of topological and non-Hermitian photonics. As compared to recent proposals and demonstrations of non-Hermitian topological lasers \cite{r23,r24}, here a synthetic imaginary gauge field \cite{r35,r36} is considered which is beneficial for laser operation: on the one hand the gauge field enables broad-area and efficient laser emission on an extended topological supermode; on the other hand, by ingenious engineering of cavity supermodes it enhances stability of laser emission. Our results provide important insights into the design of non-Hermitian laser cavities and are expected to stimulate further experimental and theoretical studies in the rapidly growing research area of integrated topological lasers and non-Hermitian photonics.

\begin{figure}[htb]
\centerline{\includegraphics[width=8.6cm]{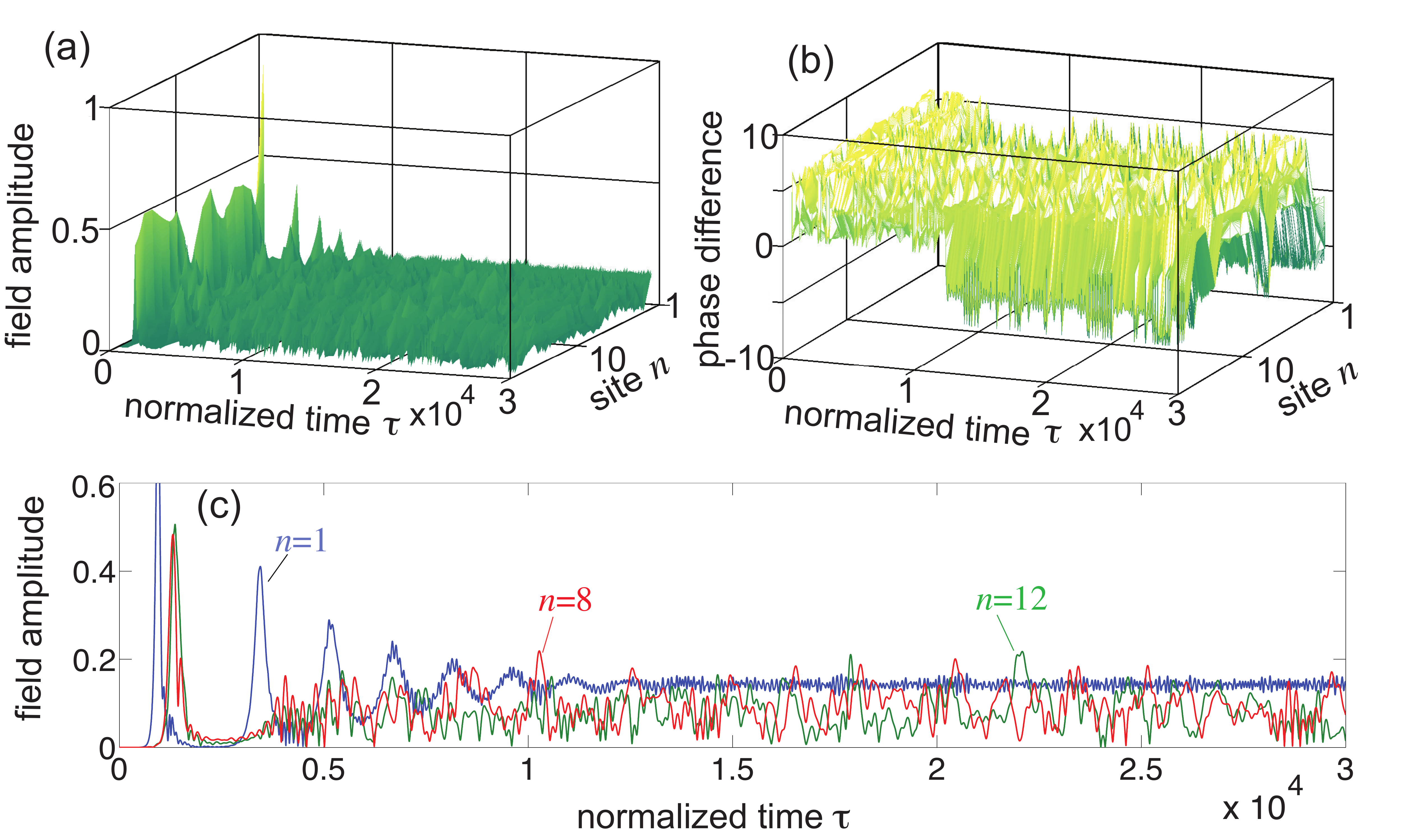}} \caption{ \small
(Color online) (a,b) Same as Fig.3, but without the synthetic imaginary gauge field ($h=0$). 
Panel (c) shows the detailed temporal behavior of modal amplitudes $|E_{n}^{(A)} (\tau)|$ in the three sites $n=1,8,12$.}
\end{figure}

\newpage

%%%%%%%%%%%%%%%%%%%%%%%%%%%%%%%
% References with full titles %
%%%%%%%%%%%%%%%%%%%%%%%%%%%%%%%

%\footnotesize
 {\bf References with full titles}\\
 \\
 \noindent
 1.  L. Lu, J.D. Joannopoulos, and M. Solja\v{c}ic, {\it Topological photonics}, Nat. Photon. {\bf 8}, 821 (2014).\\
 2. M.C. Rechtsman, J.M. Zeuner, Y. Plotnik, Y. Lumer, D. Podolsky, F. Dreisow, S. Nolte, M. Segev, and A. Szameit, {\it Photonic Floquet topological insulators}, Nature {\bf 496}, 196 (2013).\\
 3. A.B. Khanikaev and G. Shvets, {\it Two-dimensional topological photonics}, Nat. Photon. {\bf 11}, 763 (2017).\\
 4. C.M. Bender, {\it $\mathcal{PT}$ symmetry in quantum physics: From a mathematical curiosity to optical experiments}, Europhys. News {\bf 47} (2), 17 (2016).\\
 5. A. Regensburger, C. Bersch, M.-A. Miri, G. Onishchukov, D.N. Christodoulides, and U. Peschel, {\it Parity-time synthetic photonic lattices}, Nature {\bf 488}, 167 (2012).\\
 6. L. Feng, R. El-Ganainy, and L. Ge, {\it Non-Hermitian photonics based on parity-time symmetry}, Nat. Photon. {\bf 11}, 752 (2017).\\
 7.  M. Aidelsburger, S. Nascimbene, and N. Goldman, {\it Artificial gauge fields in materials and engineered systems}, arXiv:1710.00851 (2017).\\
 8. H. Hodaei, M.-A. Miri, M. Heinrich, D.N. Christodoulides, and M. Khajavikhan, {\it Parity-time-symmetric microring lasers}, Science {\bf 346}, 975 (2014).\\
 9. L. Feng, Z.J. Wong, R.-M. Ma, Y. Wang, and X. Zhang, {\it Single-mode laser by parity-time symmetry breaking}, Science {\bf 346}, 972 (2014).\\
 10. B. Peng, S.K. \"{O}zdemir, M. Liertzer, W. Chen, J. Kramer, H. Yilmaz, J. Wiersig, S. Rotter, and L. Yang, {\it Chiral modes and directional lasing at exceptional points}, Proc. Nat. Acad. Sci. {\bf 113},  6845 (2016).\\
11. P. Miao, Z. Zhang, J. Sun, W. Walasik, S. Longhi, N.M. Litchinitser, and L. Feng, {\it Orbital angular momentum microlaser}, Science {\bf 353}, 464 (2016).\\  
12. S. Longhi and L. Feng, {\it Unidirectional lasing in semiconductor microring lasers at an exceptional point}, Photon. Res. {\bf 5}, B1 (2017).\\
13. Z.J. Wong, Y.L. Xu, J. Kim, K. O$^{\prime}$Brien, Y. Wang, L. Feng, and X. Zhang, {\it Lasing and anti-lasing in a single cavity},
Nat. Photon. {\bf 10}, 796 (2016).\\
14. L. Pilozzi and C. Conti, {\it Topological lasing in resonant photonic structures}, Phys. Rev. B {\bf 93}, 195317 (2016).\\
15. B. Bahari, A. Ndao, F. Vallini, A. El Amili, Y. Fainman, and B. Kante, {\it Nonreciprocal lasing in topological cavities of arbitrary geometries}, Science {\bf 358},  636 (2017).\\
16. H. Schomerus, {\it Topologically protected midgap states in complex photonic lattices}, Opt. Lett. {\bf 38}, 1912 (2013).\\
17. S. Malzard, C. Poli, and H. Schomerus, {\it Topologically Protected Defect States in Open Photonic Systems with Non-Hermitian Charge-Conjugation and Parity-Time Symmetry}, Phys. Rev. Lett. {\bf 115}, 200402 (2015).\\
18. C. Poli, M. Bellec, U. Kuhl, F. Mortessagne, and H. Schomerus, {\it Selective enhancement of topologically induced interface states in a dielectric resonator chain},
Nat. Commun. {\bf 6}, 6710 (2015).\\
19. J.M. Zeuner, M.C. Rechtsman, Y. Plotnik, Y. Lumer, S. Nolte, M.S. Rudner, M. Segev, and A. Szameit, {\it Observation of a Topological Transition in the Bulk of a Non-Hermitian System}, Phys. Rev. Lett. {\bf 115}, 040402 (2015).\\
20. H. Zhao, S. Longhi, and L. Feng, {\it Robust light state by quantum phase transition in non-Hermitian optical materials}, Sci. Rep. {\bf 5}, 17022 (2015).\\
21. D. Leykam, K. Y. Bliokh, C. Huang, Y. D. Chong, and F. Nori, {\it Edge Modes, Degeneracies, and Topological Numbers in Non-Hermitian Systems}, Phys. Rev. Lett. {\bf 118}, 040401 (2017).\\
22. S. Weimann, M. Kremer, Y. Plotnik, Y. Lumer, S. Nolte, K.G. Makris, M. Segev, M. C. Rechtsman, and A. Szameit, {\it Topologically protected bound states in photonic parity-time-symmetric crystals}, Nat. Mater. {\bf 16}, 433 (2017).\\
23. H. Zhao, P. Miao, M.H. Teimourpour, S. Malzard, R. El-Ganainy, H. Schomerus, and L. Feng, {\it Topological Hybrid Silicon Microlasers},  	arXiv:1709.02747 (2017).\\
24. M. Parto, S. Wittek, H. Hodaei, G. Harari,  M.A. Bandres, J. Ren, M.C. Rechtsman, M. Segev, D.N. Christodoulides, and M. Khajavikhan, {\it Complex Edge-State Phase Tansitions in 1D Topological Laser Arrays}, arXiv:1709.00523v1 (2017).\\
25. W.-P. Su, J. R. Schrieffer, and A. J. Heeger, {\it Solitons in Polyacetylene}, Phys. Rev. Lett. {\bf 42}, 1698 (1979).\\
26. A. F. Glova, {\it Phase locking of optically coupled lasers}, Quantum Electron. {\bf 33}, 283 (2003).\\
27. T.Y. Fan, {\it Laser Beam Combining for High-Power, High-Radiance Sources}, IEEE J. Sel. Top. Quantum Electron. {\bf 11}, 567 (2005).\\
28. E. Kapon, J. Katz, and A. Yariv, {\it Supermode analysis of phase-locked arrays of semiconductor lasers}, Opt. Lett. {\bf 10}, 125 (1984).\\
29. H.G. Winful and S.S. Wang, {\it Stability of phase locking in coupled semiconductor laser arrays}, Appl. Phys. Lett. {\bf 53}, 1894 (1988).\\
30. H.G. Winful and L. Rahman, {\it Synchronized Chaos and Spatiotemporal Chaos in Arrays of Coupled Lasers}, Phys. Rev. Lett. {\bf 65}, 1575 1990).\\
31. R.d. Li and T. Erneux, {\it Preferential instability in arrays of coupled lasers}, Phys. Rev. A {\bf 46}, 4252 (1992).\\
32. C. P. Lindsey, E. Kapon, J. Katz, S. Margalit, and A. Yariv, {\it Single contact tailored gain phased array of semiconductor lasers}, Appl. Phys. Lett. {\bf 45}, 722 (1984).\\
33. J. Katz, S. Margalit, and A. Yariv, {\it Diffraction coupled phase-locked semiconductor laser array}, Appl. Phys. Lett. {\bf 42}, 554 (1983).\\
34. J. R. Leger, M. L. Scott and W. B. Veldkamp, {\it Coherent addition of AlGaAs lasers using microlenses and diffractive coupling}, Appl. Phys. Lett. {\bf 52}, 1771 (1988).\\
35. T.-Y. Kao, J.L. Reno, and Q. Hu, {\it Phase-locked laser arrays through global antenna mutual coupling}, Nat. Photon. {\bf 10}, 541 (2016).\\
36. N. Hatano and D.R. Nelson, {\it Localization Transitions in Non-Hermitian Quantum Mechanics}, Phys. Rev. Lett. {\bf 77}, 570 (1996).\\
37. S. Longhi, D. Gatti, and G. Della Valle, {\it Robust light transport in non-Hermitian photonic lattices}, Sci. Rep. {\bf 5}, 13376 (2015).\\
38. S. Longhi, {\it Tight-binding lattices with an oscillating imaginary gauge field}, Phys. Rev. A {\bf 94}, 022102 (2016).\\

%L. Jin, {\it Topological phases and edge states in a non-Hermitian trimerized optical lattice}, Phys. Rev. A {\bf } (2017).\\

\end{document}